# Epidemic spreading in an expanded parameter space: the supercritical scaling laws and subcritical metastable phases

Gaetano Campi[1,2], Antonio Valletta[3], Andrea Perali[1,4], Augusto Marcelli[1,5], Antonio Bianconi[1,2,6]

[1]Rome International Centre Materials Science Superstripes RICMASS via dei Sabelli 119A, 00185 Rome, Italy; antonio.bianconi@ricmass.eu
[2]Institute of Crystallography, CNR, via Salaria Km 29. 300, Monterotondo Stazione, Roma I-00016, Italy; gaetano.campi@ic.cnr.it
[3]Institute for Microelectronics and Microsystems, IMM, Consiglio Nazionale delle Ricerche CNR Via del Fosso del Cavaliere 100, 00133 Roma, Italy
[4]School of Pharmacy, Physics Unit, University of Camerino, 62032 Camerino (MC), Italy. andrea.perali@unicam.it
[5]INFN - Laboratori Nazionali di Frascati, 00044 Frascati (RM), Italy; augusto.marcelli@lnf.infn.it
[6]National Research Nuclear University MEPhI (Moscow Engineering Physics Institute), 115409 Moscow, Russia

**Abstract**

So far most of the analysis of Coronavirus 2020 epidemic data has been focusing on a short-time window and consequently a quantitative test of statistical physical laws of Coronavirus Epidemics with Containment Measures (CEwCM) is currently lacking. Here we report a quantitative analysis of CEwCM over 230 days, covering the full-time lapse of the first epidemic wave. We use a 3D phase diagram tracking the simultaneous evolution of the doubling time $T_d(t)$ and reproductive number $R_t(t)$ showing that this expanded parameter space is needed for biological physics of CEwCP. We have verified that in the supercritical [$R_t(t)>1$, $T_d(t)<40$ days] regime i) the curve $Z(t)$ of total infected cases, follows the growth rate called Ostwald law; ii) the doubling time follows the exponential law $T_d(t)=A\ e^{(t-t_0)/s}$ as a function of time and iii) the power law $T_d(t)=C\ (R_t(t)-1)^{-\nu}$ is verified with the exponent $\nu$ depending on the definition of $R_t(t)$. The log-log plots $T_d(t)$ versus ($R_t-1$) of the second 2020 epidemic wave unveils in the subcritical regime [$T_d(t)>100$ days] arrested metastable phases with $R_t>1$ where $T_d(t)$ was kept constant followed by its explosion and its containment following the same power law as in the first wave. These results provide new tools for early prediction of epidemic evolution.

## 1. Introduction

While the mean-field theory of intrinsic dynamics of uncontrolled epidemics is well known[1,2], nowadays the scientific discussion is focusing on the dynamics of *Epidemics with Containment Measures*[2-5], addressing the role of extrinsic effects due to the spatio-temporal evolution of contact networks[6-9]. In this work we verify proposed mathematical laws driving Coronavirus 2020 Epidemics with Containment Measures (called here CEwCM). Different epidemiology protocols such as *Lockdown, case Finding, mobile Tracing* (LFT)[10-16] and *Lockdown Stop and Go* (LSG)[17-19] has been applied by different countries.

While previous works[20-30] have analyzed short-time intervals of the CEwCM, we analyze here the full-time window i.e., 230 days of the first Covid-19 wave and the second 2020 wave in Italy. We focus on data analysis of three representative countries: i) South Korea, which applied the LFT policy compared with ii) Italy and iii) United States of America which applied the LSG policy with strict and loose rules respectively. We have tested key physical laws of the time evolution[24-30] of the CEwCM using a new 3D expanded parameter space, $T_d(t, R_t)$: where $T_d(t)$ and $R_t(t)$ are the time-dependent doubling and reproductive number.

In CEwCM time evolution three main regimes are clearly identified: *supercritical, critical,* and *subcritical phase.* In the *supercritical* phase the extrinsic effects control the characteristic time s in the exponential law of the time-dependent doubling time $T_d(t)=Ae^{(t-t0)/s}$. We have verified here that in the first wave this phase is characterized by the $T_d(t)=C(R_t(t)-1)^{-v}$ power law function of the variable doubling time $T_d(t)$ versus the reproductive number $R_t(t)$.

A key result of this work is the use of the log-log plots of $T_d(t)$ versus $(R_t(t)-1)$ to understand the time evolution of coronavirus first wave which is used to characterize the time evolution of epidemics in different countries. The results provide a quantitative comparison of the Covid-19 first wave evolution resulting from different choices of containment policies and the evolution of the second wave in Italy compared with South Korea.

**2. Results and Discussion**

The data for each country have been taken from the recognized public data base *OurWorldInData*[31]. We have extracted, first, the time-dependent doubling time $T_d(t)$ from the curve of total infected cases, $Z(t)$, and, second, the time-dependent reproductive number $R_t(t)$ from the curve of active infected cases, $X(t)$, using the methodological definition provided by the Koch Institute[31]. The qualitative results of this approach have been verified by producing the log-log plots of $T_d(t)$ versus $(R_e(t)-1)$ where the effective reproductive number $R_e(t)$ and $T_d(t)$ have been extracted from joint $Z(t)$ and $X(t)$ curves by using the inverted SIR model.

*2.1 The basic doubling time $T_{d0}$ and basic reproductive number $R_0$.*

Fig. 1a shows a pictorial view of the viral epidemic spreading starting with the uncontrolled epidemic in the early days of the outbreak with a *basic reproductive number* $R_0=2$ and the basic doubling time $T_{d0}=2$ days (which separates successive $n^{th}$ generations with $2^n$ infected persons).

In the early days the cumulative curve of the total number of cases $Z(t)$ increases exponentially with a characteristic rate $\alpha$

$$Z(t) = Z(0)e^{\alpha t} = Z(0) \, 2^{\frac{t}{T_{d0}}} \qquad (1)$$

therefore, the *basic doubling time*

$$T_{d0} = \frac{\log(2)}{\alpha}. \qquad (2)$$

has been quickly extracted by several groups showing that it is in the range $2<T_{d0}<2.8$ days with $2<R_0<3$, as reported in several works.[19]

*2.2 Time evolution law of the time-dependent doubling time $T_d(t)$ and reproduction number $R_t(t)$ in the 3D phase diagram $T_d(t,R_t)$*

As the epidemic goes forward, the total number of cases $Z(t)$ will include both the active cases and the removed (recovered) individuals. The time evolution of the epidemics modified by extrinsic effects of the containment measures is tracked by the time-dependent doubling time, $T_d(t)$. The time-dependent $T_d(t)$ of Coronavirus 2020 epidemics is obtained by fitting the cumulative curve $Z(t)$ over a five days period centered at t±Δt with Δt=2 days. $T_d(t)$ has been evaluated by taking the time derivative of the cumulative curve $Z(t)$ over five days of the logarithm of the cumulative infection curve $\frac{d}{dt}\log(Z(t)) = \alpha(t) = \frac{\log(2)}{T_d(t)}$ that leads to

$$T_d(t) = \frac{\log(2)}{\frac{d}{dt}\log(Z(t))}. \qquad (3)$$

The time-dependent variable reproductive number $R_t$ can be measured by different methods[2-7,32-34]. We have first used the model independent procedure proposed by the Koch Institute[33], where $R_t$ is the ratio of the actually infected individuals X(t) at time t divided by the actually infected individuals at the time (t − Δt):

$$R_t(t) = \frac{X(t)}{X(t-\Delta t)} \qquad (4)$$

The relationship between the time-dependent doubling time $T_d$ and the reproductive number $R_t$ as defined by the Koch Institute[33] can be derived by considering that

$$Z(t+\Delta t) = 2^{\frac{\Delta t}{T_d}} Z(t) = \exp\left(\log(2)\frac{\Delta t}{T_d}\right) Z(t) \cong \left(1 + \log(2)\frac{\Delta t}{T_d}\right) Z(t) \qquad (5)$$

where the equality holds true if $\Delta t < T_d$.

Hence, we find that

$$R_t(t) - 1 = \log(2)\frac{\Delta t}{T_d(t)} \qquad (6)$$

and

$$T_d(t) = C_K (R_t(t) - 1)^{-\nu} \qquad (7)$$

with

$C_K = \Delta t \, \log(2)$ and the exponent $\nu = 1$ under the hypothesis that $\Delta t$ is small.

We have used in this work $\Delta t = 5$ days, which gives a values of $R_t$ in qualitative agreement with other methods.[34] Fitting the data where the empirical $R_t$ value calculated by the Koch method using the time interval, Δt=5 days, we have found the exponent $\nu \approx 0.7 \pm 0.05$ in the three studied countries.

Fig. 1b shows the expanded three-dimensional parameter space $T_d(t',R_t)$ proposed in this work. The joint plot of the calculated $T_d(t)$, $T_d(R_t)$ and $R_t(t)$ functions provides an exhaustive description of the epidemic spreading during the time interval $t'=t-t_0=230$ days, where $t_0$ is the day onset of the first wave of Covid-19 outbreak in the three studied countries. The gray slab indicates the critical regime where $40<T_d(t)<100$ days and $R_t(t^*) \sim 1$. Around the time $t^*$ the curve of the active infected cases $X(t)$ of the Covid-19 epidemic wave reaches the maximum of a dome.

The gray critical zone separates the *supercritical* regime from the *subcritical* regime.

The *supercritical* regimes occurs in the lower part of Fig. 1b characterized by a shorter doubling time $T_d(t)<T_d(t^*)=40$ days, for $t'<t^*$ and a larger reproductive number $R_t(t)>1$ which occurs where the curve of the active infected cases $X(t)$ is growing.

The *subcritical* regime occurs in the upper part of Fig. 1b where $T_d(t)>100$ days and $R_t(t)<1$, i.e., where the curve $X(t)$ is decreasing. Fig. 1c shows the projection of the 3D plot in the $T_d$-$R_t$ plane for the three considered countries. This figure allows us to verify that in the supercritical regime (orange area) $T_d(R_t)$ follows the universal power law of Eq. 8 characterized by the divergence of $T_d(R_t)$ while approaching $R_t=1$ (thick dotted line). In the subcritical phase we see a random distribution of the data pairs ($T_d$, $R_t$) indicating the incoherent arrested phase in the green area. Fig. 1d shows both $T_d(t)$ and $R_t(t)$ vs. the time $t'=t-t_0$ for each country. The critical phase is indicated by the full grey slab as in the 3D plot of Fig. 1b, which corresponds to the range [40-100] days of the doubling time $T_d(t)$ and the reproductive number in the range $1<R_t(t)<1.1$.

The *supercritical* regime is indicated by the yellow areas, ranging from the outbreak threshold time $t_0$ and the day $t^*$, where $T_d(t^*)$ reaches the value of 40 days. The straight line $T_d(t)$ in the semi-logarithm scale in the yellow region shows that in the *supercritical* regime, the doubling time $T_d(t)$ follows the universal exponential law[10,27]

$$T_d(t)=Ae^{(t-t_0)/s} \qquad (8)$$

where $t_0$ is the time of the onset of the outbreak and s (called the s-factor) is the characteristic time of the applied containment policy[10,27]. The upper panel shows that the time lapse of the supercritical phase $t^*-t_0 =24$ days for the yellow region of South Korea is related with the small s-factor (6-7 days) which are both a factor 2.5 shorter than the s-factor of Italy and USA where $60<(t^*-t_0)<70$ days, which imply a time duration of the first wave lockdown a factor 2-3 longer.

The arrested phase occurs in the light blue areas where $T_d(t)$ becomes larger than 100 and $R_t(t)$ drops below 1 (see green areas in Figure 1d).

Italy reached a well defined arrested phase extending in the green area due to the enforced strict rules of the imposed lockdown. On the contrary, the plot shows that in USA, where loose rules for the lockdown have been applied, the transition to the subcritical regime has been stopped, and the country remained for a very long time in the critical phase. This approach shows that the arrested phase in South Korea remained close to the critical phase and faced a second Covid-19 very short wave indicated by the yellow area. We underline that the second wave in South Korea lasted only 20 days and with a doubling time longer than 40 days. The shape of the second wave in South Korea clearly shows the efficiency of the applied contact tracing to keep the spreading under control, to reduce the lockdown to a minimum time and to reduce the number of fatalities.

*2.3 Infection and removal time from inverted SIR model: the $R_e(t)$ time-dependent reproduction time in the $T_d(t, R_e)$ phase diagram*

We move now to describe the $T_d(t',R_e)$ phase diagram, where the effective reproduction time, namely $R_e(t)$, is extracted by the inverted SIR theory[25,26].

In the framework of the SIR mean-field theory, the individuals in a population are classified as *Susceptible* (if they haven't yet developed the infection), *Infected* (if they currently carry the infection) and *Removed* (if they have recovered from the infection or dead or they have been vaccinated or confined by strict rules, see for instance the case of the Covid-hotels organized in some countries or regions). The rate of change of these populations is calculated by introducing two characteristic times: the infection time ($\tau_i$, the mean time between two infections made by an infected individual) and the removal time ($\tau_r$, the mean time needed for an individual that has been infected to recover from the disease or to die).

While in the classical epidemiology, for an uncontrolled epidemic spreading the infection time and the removal time have constant values characteristic of the outbreak, in a controlled epidemic spreading after the introduction of containment measures both the infection and removal times are time-dependent. If we indicate with N the size of the population, then the system of ordinary differential equations governing S, X and Y the number of *Susceptible*, *Infected*, and *Removed* individuals, respectively can be written as:

$$\begin{cases} \frac{dS}{dt} = -\frac{X(t)}{\tau_i(t)}\frac{S(t)}{N} \\ \frac{dX}{dt} = \frac{X(t)}{\tau_i(t)}\frac{S(t)}{N} - \frac{X(t)}{\tau_r(t)} \\ \frac{dY}{dt} = \frac{X(t)}{\tau_r(t)} \end{cases} \qquad (9)$$

The effective reproduction number obtained by inverting the SIR equations is called $R_e(t)$ which is defines as

$$R_e(t) = \frac{\tau_r(t)}{\tau_i(t)} \qquad (10)$$

under the hypothesis that the relative change in the number of *Susceptibles* is negligible, therefore S can be considered almost constant and equal to N. Hence the second equation simplify to:

$$\frac{dX}{dt} = \left(\frac{1}{\tau_i(t)} - \frac{1}{\tau_r(t)}\right) X(t) \quad (11)$$

pointing out that the rate of changes of the *Infected* is related to the ratio of $\tau_r$ to $\tau_i$ that is called here the effective reproduction number $R_e(t)$.

The values of $R_e(t)$ for the three countries calculated with Eq. 11 are shown in Fig. 2.

The amount of currently infected or active positive cases (I) and the total number Z = X + Y of individuals that are affected, at any time, by the infection are available on the data base. The rate equation for Z is

$$\frac{dZ}{dt} = \frac{X(t)}{\tau_i(t)} \frac{S(t)}{N} \approx \frac{X(t)}{\tau_i(t)} \quad (12)$$

hence, starting from the pandemic data for X and Z and using the approximation S(t) ≈ N, it is possible to *invert the SIR model to* evaluate $\tau_r$ and $\tau_i$

$$\begin{cases} \tau_r(t) = \frac{X(t)}{\frac{dZ}{dt} - \frac{dX}{dt}} \\ \tau_i(t) = \frac{X(t)}{\frac{dZ}{dt}} \end{cases} \quad (13)$$

Using Eq. 13 we have extracted the *infection time,* $\tau_i$, and *removal time,* $\tau_r$, shown in Figure 2 in three selected countries. Moreover, we have calculated the amount of the actually infected, the removed and the total cases by using the extracted $\tau_i$ and $\tau_r$ values and compared the results with the epidemic data. The curves quantitatively agree thus validating the consistency of the procedure.

Figure 2a shows that the transition from the supercritical to the subcritical regime is driven by the joint increase of $\tau_i$ and decrease of $\tau_r$. The opposite occurs when the subcritical to the supercritical transition happens in metastable phases at end of the first wave and the threshold of the second wave.

In Fig. 2b and Fig. 2c a similar analysis is performed on the epidemic data of Italy and USA respectively. The comparison among countries shown in Figure 2a, 2b and 1c indicates different efforts in testing and tracing.

The cumulative curve Z(t) of the total number of cases of the epidemics in the supercritical phase, slows down approaching the critical regime, where it has been fitted by the complex Ostwald growth law[10,17,35-37]

$$Z(t - t_0) = C\{1 - e^{-(t-t_0)/\tau}\} \cdot (t - t_0)^\gamma \quad (14)$$

which is a mixed exponential and power-law behavior determined by nucleation of different phases and ordering phenomena in complex multiphase systems out of equilibrium[35-37]. The onset of this behavior, obtained by best curve fitting, is reported by the dashed lines in the central panels of Fig. 2.

The time-dependent doubling time given by

$$T_d(t) = \frac{\log(2)}{\frac{d}{dt}\log(Z(t))} = \frac{\log(2)}{\frac{\frac{d}{dt}Z(t)}{Z(t)}} \quad (15)$$

can be obtained from the equation for $\tau_i(t)$. In fact, from the Eq. (12) we know that $\frac{d}{dt}Z(t) = \frac{X(t)}{\tau_i(t)}$ and, in the framework of the SIR model, we obtain the following expression for $T_d$

$$T_d(t) = \log(2)\, \tau_i(t)\, \frac{Z(t)}{X(t)} \quad (16)$$

$$T_d(t) = \log(2)\, \tau_r(t)\, \frac{Z(t)}{X(t)}\, \frac{1}{R_e(t)} \quad (17)$$

The doubling time $T_d(t)$ calculated using Eq.17, has been plotted in Figure 2 with the effective reproduction number $R_e(t)$, computed within the SIR model. The $T_d(t)$ curve estimated with the SIR model agrees with the $T_d(t)$ curve estimated from the epidemic data (shown in the upper panel of Figure 2d).

At variance the $R_e(t)$ curve exhibits only a qualitative agreement with the corresponding $R_t(t)$ curve estimated with the Koch Institute method shown in Fig.1. Nevertheless, both methods agree on the time where the critical regime occurs, i.e., when both $R_t(t)$ and $R_e(t)$ are equal to one.

*2.4 The universal power law relation between time-dependent doubling time, $T_d(t)$, and reproduction number, $R_t(t)$, in the supercritical phase: prediction of the onset of new epidemic waves.*

Log-log plot of the doubling time $T_d$ as a function of $R_t-1$ and $R_e-1$ in the three countries are compared in the upper and lower panels of Fig. 3. The $R_t$ values are higher than $R_e$ calculated with inverted SIR method. although the difference, the $T_d(t)$ vs $R_t(t)-1$ behavior in the two approach appears qualitatively similar. We get the critical exponent $\nu=1$, using $R_e-1$ extracted by the mean-field inverted SIR model which is different from the exponent $\nu=0.7$ obtained using Rt defined by the Koch method.

The exponent $\nu=1$ was predicted for the Koch $R_t$ definition in the limit $\Delta t \to 0$ by Eq.7 but this limit cannot be reached since the time interval cannot be smaller than 3 days. We have estimated $R_t$ with the Koch Institute method[33] using different time intervals $\Delta t = 3, 5, 7, 9, 11,$ and 21 days. Afterwards we have fitted the data $T_d(R_t)$ using Eq. 7 in the three selected countries using $R_t$ measured with the different time intervals $\Delta t$. We have found that the exponent $\nu$ decreases by increasing the time interval $\Delta t$ following the stretched exponential function $\nu = e^{-(\Delta t/\tau)^\beta}$ with $\tau=24.7$ days and $\beta=0.76$. Therefore, as predicted by Eq.7, we have found $\nu \to 1$ for $\Delta t \to 0$ that is in agreement with the mathematical limit for the exponent extracted by the Koch method for the time interval $\Delta t \to 0$.

This result is particularly relevant to predict the onset of new epidemic waves. The onset of the supercritical phase of successive wave will be easily detected in the [$T_d$, $R_t$-1] or [$T_d$, $R_e$-1] diagram when dots move towards the yellow area, where the supercritical regime will be following the universal power law of Eq. (7).

Fig.3 shows that in USA the epidemics never entered in the arrested phase and the it has been fluctuating in the critical phase with an increasing (black dots) and decreasing (red dots) doubling time along with the predictions of the power law.

The rapidly aborted second wave in South Korea is well depicted by the green dots forming the loop in Fig.3, where the containment policy succeeded to push up the doubling-time avoiding its following down toward the universal power law growth rate of the first Covid-19 wave.

*2.5 Metastable phases and the second wave in Italy*

In Figure 3 the phase diagram of epidemic spreading in Italy during the first wave shows the grey strip corresponding to the *critical* regime occurring for $T_d(t)$ between 40 and 100 days above the *supercritical* orange area, and the arrested *subcritical* regime, characterized by for $T_d(t)$>100 days. In this arrested *subcritical* phase the $T_d(t)$ vs $R_t(t)$ values show a non-analytical disordered distribution (see also Figure 1c), while in the supercritical regime it is described by the analytical universal power law of Eq. (7). The green dots in the $T_d$ vs. $R_t$ plots in Fig. 3 show that the arrested phase occurs also for cases where $T_d(t)$>100 although $R_t(t)$>1. When put on a lattice, the SIR model depends on the geometry of contacts, and is in the same universality class as ordinary percolation. $R_t$ is essentially the percolation threshold for the SIR mean-field model, in fact infection can grow without bound where $R_t$ is greater than 1 on a Erdos-Reny network or a Bethe lattice[38], while there is no percolation for $R_t$ less than 1. However, making the network composed of closed loops which attenuate the probability of an epidemic, in different complex geometries and in presence of long range interactions the percolating epidemic threshold could be larger than 1 (in fact it can be as high as 1.2).[39]

The plot of the Italian case unveils the unexpected metastable phases in the arrested regime with 1.06<$R_t(t)$<1.2 with a decreasing reproduction number at constant doubling-time ($R_t(t)$ decreases while $T_d(t)$ remains constant) as indicated by horizontal arrows in the curve (green dots) of the ($T_d$ vs $R_t$) plots.

After the flat metastable state in the subcritical phase at the end of the first wave on October 7, 2020 the doubling time decreased toward the supercritical regime at the end of the zig-zag behavior where the arrows indicate the time direction. In Fig. 3 the data of South Korea at the end of the flat regime show that the efficient mobile contact tracing elevated the doubling time. On the contrary in Italy the poor contact tracing method enforced in Italy caused the decrease of the doubling time with the onset of supercritical regime of the second Covid-19 epidemic wave.

The evolution of the second epidemic wave in Italy between Sept 1st and Dec 13th, is plotted in Fig. 4. The data in panel b) of Fig. 3 and the data in Fig 4 overlap in the time period Sept 1st (245 DOY) and Oct 7 (281 DOY) where we have observed the metastable arrested phase discussed above which can be considered either as the end of the first wave as well as the onset of the second wave. Panel a) of Fig. 4 shows the lapse of time where the curve of active cases has shown its maximum increasing rate due to the second wave in Italy. In the three panels if it is possible to see that from DOY 275 (Oct 1st) to DOY=300 (Oct 26) the reproductive number increased from 1.05 to 1.4 while the doubling-time decreased from $T_d$=102 days in the subcritical regime to $T_d$=19 days in the supercritical regime. The rate of the growth rate of active cases started to decrease only after DOY=300 (Oct 26). It is remarkable that in the time period 300<DOY< 330 days, when some strict containment rules have been enforced, the CEwCM developed again as in the same regime of the first wave following both Eq. 7 $T_d(t) = C_K (R_t(t) - 1)^{-\nu}$ and Eq. 8 $T_d(t)=Ae^{(t-t_0)/s}$ where the s factor is about two times larger than in the South Korea second wave shown in Fig. 3.

**Conclusions**

The results of this work provide an original quantitative approach for understanding the time evolution of the Covid-19 pandemic. We show that it is necessary to expand the parameter space, monitoring the evolution of the pair of relevant variables ($T_d$, $R_t$). By expanding the parameter space it became possible to analyze in a more precise and complete manner the data of the epidemics, probing and tuning at the same time containment measures. This work sheds light and provides new quantitative experimental tools for the quantitative statistical physics of this Covid-19 pandemic, but certainly also to face future epidemic events thanks to its predicting power.

The results of our work can be summarized as follows: the joint analysis of the *doubling time* $T_d(t)$, i.e., the time it takes for the number of infected individuals to double in value extracted from the cumulative curves of total (infected plus removed) cases, combined with the reproductive number $R_t(t)$, i.e., the average number of infected persons by a single positive case, extracted from the cumulative curves of the number of active infected cases, provide complementary information on the efficiency of the applied containment policies. Therefore, the proposed approach could be used to dynamically control and to improve the effects of mitigation policies.

The quantitative data analysis here presented has been able to test the correlated function of $T_d(t)$ vs. $R_t(t)$. The main result of this work is that the time evolution of CEwCM spreading in the supercritical regime is characterized by the substantial growth of the doubling time $T_d(t)$ for $R_t(t)$>1 and $T_d(t)<40$ days described by the power law curve $T_d = C(R_t-1)^{-\nu}$. The second key result is the identification that the $T_d(t)$ values for which the Covid-19 epidemic reaches the critical regime or the plateau at the transition between the subcritical and the supercritical phase fall in the interval 40-100 days and $Rt \approx 1$. Finally, we have clearly identified the metastable arrested phase of the diffusion phenomenon with $R_t$ ($1.02<R_t<1.16$) combined with a doubling time larger than 40 days which are precursors of the onset of a second wave, result that can be adopted as an early warning of new waves.

**Figures and Figure captions**

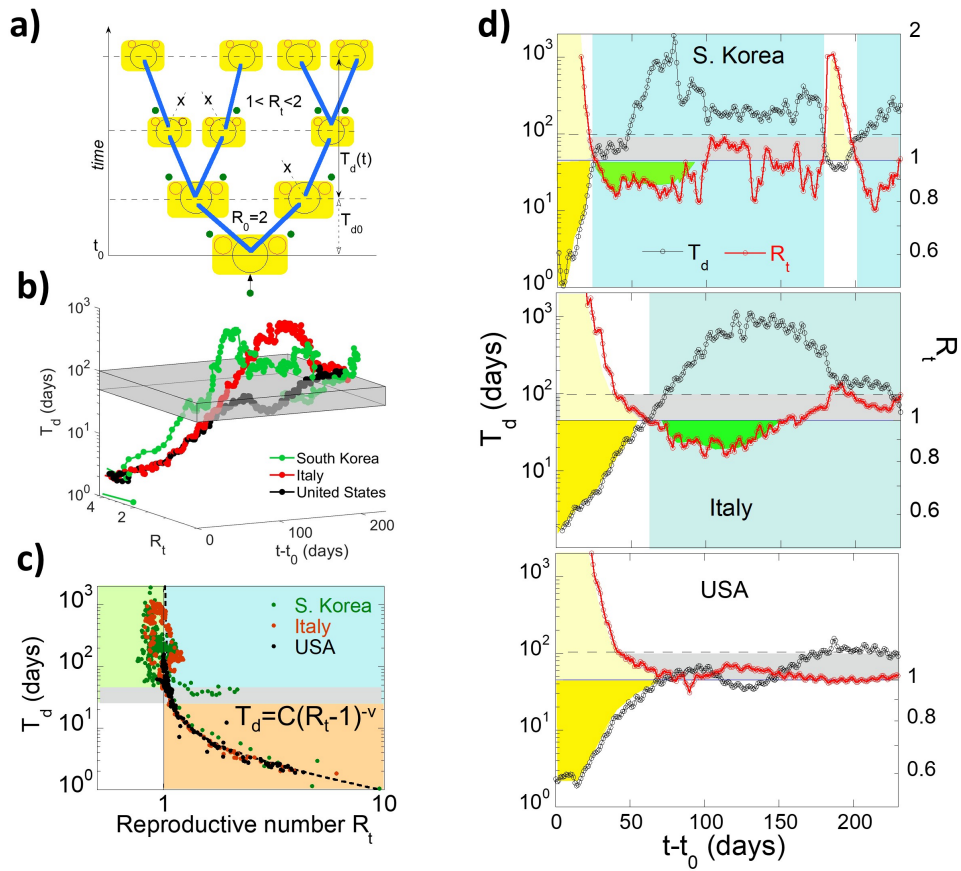

**Figure 1 (a)** Pictorial view of the epidemic growth starting with basic reproductive number, $R_0$, and basic doubling time, $T_{d0}$. Containment measures in the supercritical phase of epidemic spreading reduce the reproductive number $1<R_t<1$ and lengthen the doubling time in the range $2<T_d<40$ days. **(b)** The 3D phase diagram where the doubling time $T_d(t',R_t)$ is plotted as a function of $t'=t-t_0$ and the effective reproductive number $R_t$ for South Korea (green), Italy (red), and USA (black). The gray space indicates the critical crossover where $40<T_d<100$ days separates the supercritical phase ($T_d<40$ days; $R_t>1$) from the arrested subcritical phase ($T_d>100$ days; $R_t<1$). **(c)** Log-log plot of doubling time $T_d$ vs $R_t$ in the three countries. In the orange supercritical regime all curves of the three countries are fitted by the function $T_d(t)=C(R_t(t)-1)^{-\nu}$. The data in the subcritical regime, green area, where $T_d>100$ days and $R_t<1$, show the incoherent disordered behavior. The critical phase is confined within the horizontal grey dashed strip where $40<T_d<100$ days. **(d)** $T_d$ and $R_t$ as a function of $t'=t-t_0$ in the three selected countries. In the supercritical regime in the yellow areas the doubling time (black curve) increases in the range $2<T_d<40$ days and the reproductive number $R_t$ decreases down to 1. In the arrested phases, indicated by the light blue areas, $T_d$ increases to a maximum and $R_t$ decreases up to a minimum and it extends up to the point where they cross again in the critical regime. The green area identifies the subcritical regime for $R_t<1$ and $T_d>100$ days where the epidemic spreading is arrested. Finally, the grey area identifies the critical regime where $1<R_t<1.1$ and $40<T_d<100$ days.

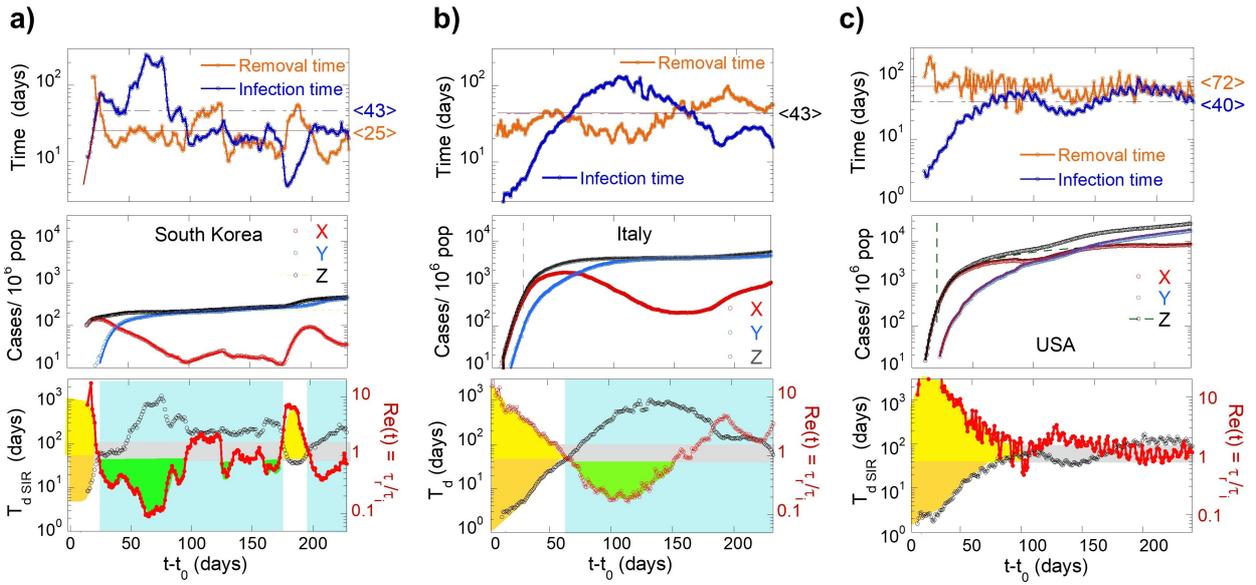

**Figure 2** The inverted SIR model applied to **(a)** South Korea, **(b)** Italy, and **(c)** USA. The upper panels show the extracted (red) removal time and (blue) infection time as a function of $t-t_0$. In the middle panels, we report the curves of cases per million population, of infected (X) (red), removed Y (blue) and total cases (Z=X+Y) (black) as a function of time. The Z(t) curve follows the Ostwald law[10,27] (dashed line) characteristic of ordering growth in heterogeneous systems[35-37]. In the lower panels we show $T_d$ and $R_t$ ($R_e$) as extracted by the inverted SIR model in analogy with Figure 2c. The shaded light blue area indicates the arrested regime in the subcritical phase separating the first and the second Covid-19 wave. In Italy this arrested phase occurs for 60<t'<230 days. In USA there is no evidence for the presence of a subcritical regime, the yellow supercritical phase is followed by the critical phase in the gray strip extending over a large time interval.

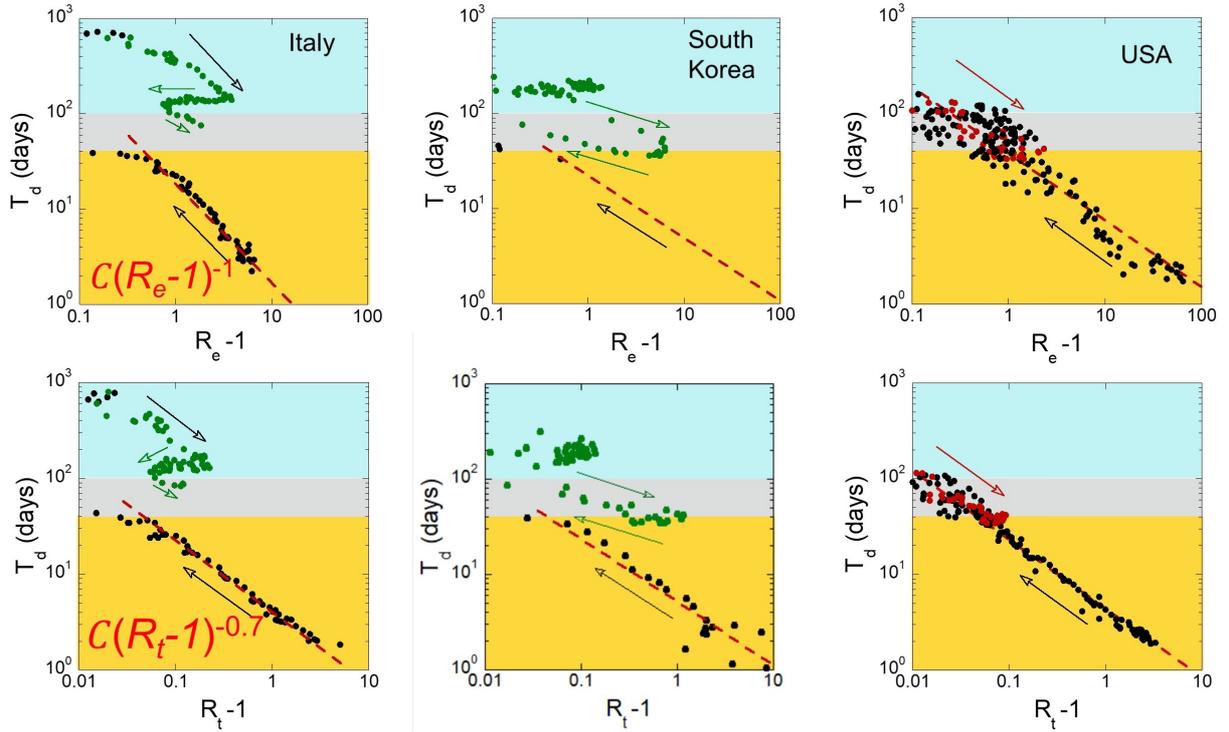

**Figure 3** Log-log plot of doubling time $T_d$ in three countries as a function of R(t)-1 extracted by the inverted SIR model (upper panels) and by the Koch Institute empirical method (lower panels). Black dots in the yellow region indicate $T_d$ in the supercritical regime which follows the universal analytical power law function $[T_d=C(R_t-1)^{-\nu}]$ of a divergent $T_d$ approaching $R_t=1$ fitted by the red dashed lines. The critical exponent $\nu$ is 1 in the inverted SIR model and 0.7 using Koch prescriptions with 5 days interval. The critical phase is described by the horizontal grey strip where $40<T_d<100$ days. The green dots in the blue light area indicate the metastable states in the subcritical regime where the epidemic spreading growth is arrested $T_d>40$ days also if $R_t>1$. While the metastable state in Italy in the subcritical (green dots) with $T_d$=constant indicates a precursor of the second wave regime in South Korea, the loop of the curve (green dots) in the critical regime (gray strip) is determined by the second wave shown in the lower panel of Fig. 2a. In USA the epidemic spreading never entered in the arrested phase and reversible oscillations ($R_t$ decreasing is inducted by red dots) ~~and~~ are observed in the critical regime.

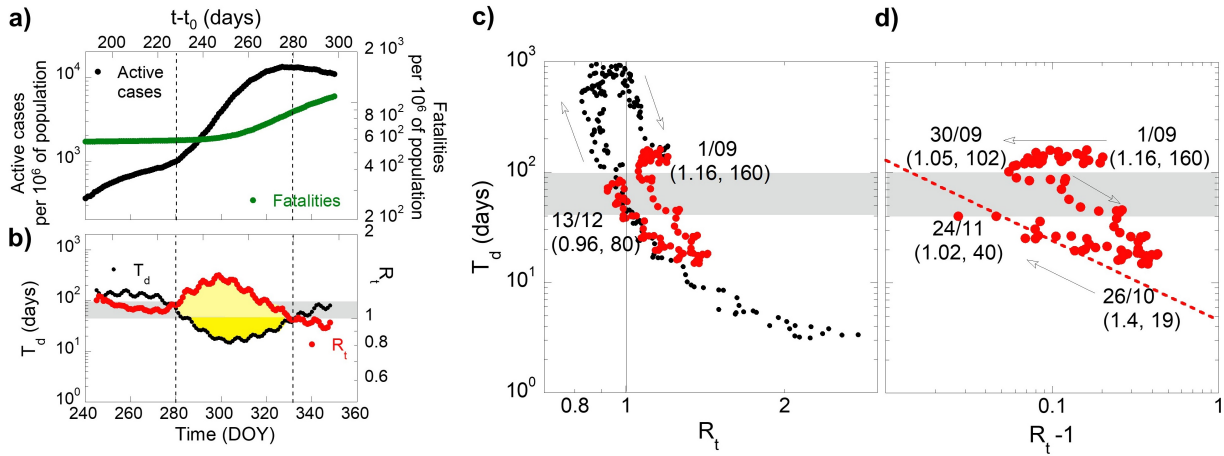

**Figure 4)** Panel **a)** the curve of the active infected cases and fatalities per million population as a function of time during the second wave in Italy. The time is measured by the day of the year 2020 in the lower scale, and the time t'=t-t$_0$ from the onset of the outbreak in Italy in the upper x scale. Panel **b)** shows $T_d(t)$ and Rt(t) for the second wave in Italy with the same scale as in Fig. 1 for the first wave in Italy. The yellow area indicates the explosive percolating epidemic regime of the second wave in Italy. Panel **c)** shows the $T_d$ versus $R_t$ plot of the second wave (red dots) compared with the first wave (black dots). Panel d) shows the log-log plot of $T_d$ versus ($R_t$-1). The red dots in the second wave shows the first uncontrolled increase of $R_t$ and decrease of $T_d$ up to its peak in the time lapse 274<DOY<300 which is followed by the coronavirus epidemics controlled by containment measures in the time range 300<DOY<330 which follows the power law given by Eq. No. 7.